\documentstyle[twoside,amssymb,12pt]{article}

\newcommand{\nc}{\newcommand}
\nc{\la}{\lambda} \nc{\alf}{\alpha}
\nc{\tht}{\theta}  \nc{\be}{\beta}  \nc{\eps}{\epsilon}
\nc{\ga}{\gamma}  \nc{\D}{\Delta}  \nc{\G}{\Gamma}  \nc{\vphi}{\varphi}
\nc{\de}{\delta} \nc{\si}{\sigma}  \nc{\ka}{\kappa}   \nc{\Si}{\Sigma}
\nc{\om}{\omega}  \nc{\Om}{\Omega}  \nc{\z}{\zeta}  \nc{\T}{\Theta}
\nc{\qq}{\quad\quad}   \nc{\nf}{\infty}   \nc{\pt}{\partial}
\nc{\dl}{\mathop{\smash{\cal L}}}  \nc{\black}{\rule{1mm}{3mm}}
\nc{\beq}{\begin{equation}}     \nc{\eeq}{\end{equation}}
\nc{\beqa}{\begin{eqnarray}}  \nc{\dst}{\displaystyle}  \nc{\sst}{\scriptstyle}
\nc{\eeqa}{\end{eqnarray}}   \nc{\nnb}{\nonumber}
\nc{\bs}{\backslash}        \nc{\mb}{\mathbb}     
\nc{\sn}{{\rm sn}\,}   \nc{\cn}{{\rm cn}\,}         \nc{\dn}{{\rm dn}\,}
\nc{\ti}{\tilde}         \nc{\wti}{\widetilde}    \nc{\wh}{\widehat}
\nc{\nin}{\noindent}     \nc{\ol}{\overline}      \nc{\ul}{\underline}
\nc{\modsim}{\mathop{\smash{\sim}}}     

\newcounter{muni}
\newenvironment{remunerate}{\begin{list}{{\rm \arabic{muni}.}}
{\usecounter{muni}
\setlength{\leftmargin}{0pt}\setlength{\itemindent}{38pt}}}{\end{list}}

\nc{\brm}{\begin{remunerate}}   \nc{\erm}{\end{remunerate}}

\newtheorem{nth}{Proposition}   \newtheorem{ndef}{Definition}
\nc{\simas}{\mathop{\smash{\sim}}}
\nc{\stg}{\mathop{\smash{*}}}   \nc{\LIM}{\mathop{\smash{\lim}}}
\nc{\st}{\mathop{\smash{\delta}}} \nc{\SUP}{\mathop{\smash{\rm sup}}}
\nc{\barr}{\begin{array}}   \nc{\earr}{\end{array}}   \nc{\dg}{\dagger}
\nc{\mtvb}{\mathversion{bold}}   \nc{\mtvn}{\mathversion{normal}}

\begin{document}
%\begin{titlepage}

\vspace{5mm}
{\noindent \bf International Conference on Difference Equations, \\ 
Special Functions and Applications, \\
Munich, 25-30 july 2005}

\vskip 1.0truecm 
\centerline{\Large\bf Heun functions versus}

\vspace{5mm}
\centerline{\Large\bf elliptic functions}

\vspace{5mm}
\centerline{\bf Galliano VALENT${}^{\;\ddagger}\ {}^*$}

\vskip 0.5truecm
\centerline{${}^{\ddagger}$ \it Laboratoire de Physique Th\'eorique et des
Hautes Energies}
\centerline{\it CNRS Unit\'e associ\'ee URA 280}
\centerline{\it 2 Place Jussieu F-75251 Paris Cedex 05 France}

\vskip 0.5truecm
\centerline{${}^*$ \it Departement de Mathematiques}
\centerline{\it Case 901  163 Avenue de Luminy}
\centerline{\it 13288 Marseille Cedex 9 France}
\nopagebreak
\vskip 0.5truecm

\begin{abstract}
\noindent We present some recent progresses on Heun functions, gathering results 
from classical analysis up to elliptic functions. We describe Picard's generalization 
of Floquet's theory for differential equations with doubly periodic coefficients 
and give the detailed forms of the level one Heun functions in terms of 
Jacobi theta functions. The finite-gap solutions give an interesting 
alternative integral representation which, at level one, is shown to 
be equivalent to their elliptic form.

\end{abstract}

%\end{titlepage}

\section{Introduction}
Heun functions \cite{He} are defined as a natural generalization of the hypergeometric 
function, to be the solutions of the Fuchsian differential equation
\beq\label{eqHn1}\left\{\barr{l}\dst 
\frac{d^2F}{dw^2}+\left(\frac{\ga}{w}-\frac{\de}{1-w}-\frac{\eps k^2}{1-k^2w}\right)
\frac{dF}{dw}+\frac{(s+\alf\be k^2w)}{w(1-w)(1-k^2 w)}F=0,\\[5mm]
\ga+\de+\eps=\alf+\be+1,\earr\right.
\eeq
with four regular singularities $0,\,1,\,1/k^2,\,\nf.$ In \cite{Ar} the fourth 
singularity is $a\equiv 1/k^2$ and the auxiliary parameter $s$ is taken 
as $q\equiv -s/k^2.$ We will follow the notation 
\beq\label{eqHn3}
Hn(k^2,s;\alf,\be,\ga,\de;w),\qq k^2\in[0,1].
\eeq
In \cite{Ar} one can find a lot of information on these functions and in the 
remaining chapters of \cite{Ron1} are gathered many results on their confluent 
limits. Due to limitation of space the former section misses several important items, 
particularly the relation between Heun functions and elliptic function theory. 
These last years this field has received new interesting developments from workers 
in condensed matter physics \cite{BES} and integrable systems \cite{Kr}. Interesting 
accounts have been given by Smirnov \cite{Sm} and by Takemura \cite{Ta}.

Our aim is to gather, for the community dedicated to the study of special functions 
and orthogonal polynomials, these new progresses relating Heun and elliptic 
functions. 

As a preliminary step we will give one further motivation for the study of Heun 
functions, coming from orthogonal polynomials and birth and death processes.

\section{From orthogonal polynomials to Heun functions}
Let us consider the three terms recurrence \cite{Ak1}
\beq\label{b1}
xP_n=b_{n-1}\,P_{n-1}+a_nP_n+b_n\,P_{n+1},\qq n\geq 1.\\[4mm]
\eeq
We will denote by $P_n$ and $Q_n$ two linearly independent solutions of this 
recurrence with initial conditions
\beq\label{b2}
P_0(x)=1,\qq P_1(x)=\frac{x-a_0}{b_0},\qq\qq Q_0(x)=0,\qq Q_1(x)=\frac 1{b_0}.\eeq
The corresponding Jacobi matrix is 
\beq\label{Jacobi}
\left(\barr{cccccc}
a_0 & b_0 & 0 & 0 & 0 & \cdots\\[4mm]
b_0 & a_1 & b_1 & 0 & 0 & \cdots\\[4mm]
0 & b_1 & a_2 & b_2 & 0 & \cdots\\[4mm]
\cdots & \cdots & \cdots & \cdots & \cdots & \cdots\\[4mm]
\cdots & \cdots & \cdots & \cdots & \cdots & \cdots \earr\right)
\eeq
If the $b_{n}>0$ and $a_{n}\in{\mb R}$ the $P_n$ (resp. the $Q_n$) will be 
orthogonal with respect to a positive probabilistic measure $\psi$ (resp $\psi^{(1)}$)
\beq\label{b3}
\int \,P_m(x)\,P_n(x)\,d\psi(x)=\de_{mn},\qq\quad 
\int \,Q_m(x)\,Q_n(x)\,d\psi^{(1)}(x)=\de_{mn},\eeq
with the moments
\beq\label{b4}
s_n=\int x^n\,d\psi(x),\qq n\geq 1,\qq s_0=1\eeq

When considering applications to birth and death processes with killing \cite{KT}, 
it is sufficient to consider the special case where
\[\barr{l}
a_n=\la_n+\mu_n+\ga_n,\qq\qq b_n=\sqrt{\la_n\mu_{n+1}},\quad n\geq 0,\\[4mm]\dst 
\pi_0=1,\qq\pi_n=\frac{\la_0\la_1\dots\la_{n-1}}{\mu_1\mu_2\cdots\mu_n},\quad n\geq 1.
\earr\]
It is convenient to introduce new polynomials $F_n(x)$ by
\beq\label{mt1}
P_n(x)=\frac{(-1)^n}{\sqrt{\pi_n}}\,F_n(x),\qq n\geq 0,\qq
\eeq
The positivity constraints 
\beq\label{pos}
\la_n>0, \qq\qq \mu_{n+1}>0\, \qq\qq n\geq 0,
\eeq 
will be assumed. From (\ref{b1}) we deduce
\beq\label{mt3}\barr{l}
-xF_n=\mu_{n+1}F_{n+1}-(\la_n+\mu_n+\ga_n)F_n+\la_{n-1}F_{n-1},\\[4mm]
 F_{-1}(x)=0,\quad F_0(x)=1.\earr
\eeq
The orthogonality relation (\ref{b3}) becomes 
\beq\label{oph1}
\int \,F_m(x)\,F_n(x)\,d\psi(x)=\pi_n\de_{mn}. 
\eeq

The particular case where $\la_n,\,\mu_n$ are quadratic is connected with 
Heun functions in the following way. Let us take
\beq\label{oph2}\barr{l}
\la_n=k^2(n+\alf)(n+\be),\qq\mu_n=n(n+\ga-1),\qq\ga_n=k'\,^2\de n\\[4mm]
\qq\alf>0,\ \be>0,\ \ga>0,\quad k^2\in(0,1),\qq k'\,^2=1-k^2,\earr
\eeq
and consider the generating function
\beq\label{oph3}
F(x;w) =\sum_{n\geq 0}\,F_n(k^2,x;\alf,\be,\ga,\de)w^n\equiv 
Hn(k^2,s;\alf,\be,\ga,\de;w),\qq s=x-\alf\be k^2.
\eeq
Routine computations show that $F(x;w)$ is the solution, analytic in a neighbourhood
of $w=0$ \footnote{In \cite{Ron1} this function is called ``local" Heun. Notice that 
we have for normalization $\,Hn(\cdots;0)=1.$}, 
of Heun's differential equation (\ref{eqHn1}). Quite remarkably, the most 
general quadratic birth and death process with linear killing produces, for its 
generating function, the most general Fuchsian equation with four regular 
singularities.

Let us conclude with the proof of:

\begin{nth} 
The Hamburger (hence the Stieltjes) moment problem corresponding to the 
polynomials $F_n(x)$ with recurrence coefficients (\ref{oph2}) is determinate.
\end{nth} 

{\bf Proof :}

\nin A theorem by Hamburger \cite{Ak1}[p. 84] states that if the series $c_n=P_n(0)^2$ 
is divergent then the Hamburger moment problem is determinate, i. e. the measure $\psi$ 
defined in (\ref{b3}) is unique. From relation (\ref{mt1}) we can write
\[\frac{c_{n+1}}{c_n}=\frac{\mu_{n+1}}{\la_n}\, \frac{F_{n+1}(0)^2}{F_n(0)^2}
=\frac 1{k^2}\,\frac{(n+1)(n+\ga)}{(n+\alf)(n+\be)}\,\frac{F_{n+1}(0)^2}{F_n(0)^2}.\]
Now the generating function $F(x;w),$ introduced in (\ref{oph3}), is 
analytic for $\,|w|<1\,$: hence its radius of convergence is one. It follows that, 
for $n\to\nf$ the previous ratio has for limit $1/k^2>1$ and the 
series $\{c_n\}$ diverges. $\qq\Box$

To have a clear view of the problems encountered in the construction of solutions of 
Heun's equation, we will introduce some terminology: we will 
call ``generic" solutions the solutions valid for 
{\em arbitrary values} of the auxiliary parameter $s$ and ``non-generic" solutions 
those which require particular values of $s.$  Let us begin with some results 
on non-generic solutions.

\[  \]

\centerline{\bf\Huge Non-generic solutions}

\[  \]

\setcounter{section}{0}
\section{Special hypergeometric cases}
The reason for the restriction $k^2\in(0,1)$ is that in the two limiting cases 
$k^2=0$ (notice that the positivity conditions (\ref{pos}) require $k^2>0$)  
and $k^2=1$ the four singular points reduce to three and therefore Heun functions 
degenerate into hypergeometric functions. Indeed we have:
\brm
\item For $k^2=0$ the parameters $\,(\alf,\,\be)\,$ become irrelevant and we have:
\beq\label{red1}\left\{\barr{l}\dst 
Hn(0,s;(\alf,\be),\ga,\de;w)=\, 
_2F_1\left(\barr{c} r_+,r_-\\ \ga\earr \ ; w\right),\\[5mm]\dst
r_{\pm}=\rho\pm\sqrt{\rho^2+s},\qq\qq  \rho=\frac 12(\ga+\de-1).\earr\right.
\eeq
This case, which would correspond to $a\to\nf$, is missing in \cite{Ar}. 
\item For $k^2=1$ we have the relation:
\beq\label{red2}\left\{\barr{l}\dst 
Hn(1,s;\alf,\be,\ga,\de;w)=
(1-w)^r\, _2F_1 \left(\barr{c} r+\alf,r+\be\\ \ga\earr \ ; w\right),\\[5mm]\dst 
r=\rho-\sqrt{\rho^2-\alf\be-s},\qq \rho=\frac 12(\ga-\alf-\be).\earr\right.
\eeq
This case is considered in \cite{Ar} but only with the extra constraint $s=-\alf\be,$ 
i. e. for $r=0.$
\erm

\section{The ``trivial" solution}
This solution corresponds to the special values $s=\alf\be=0.$ In this case 
Heun's differential operator is factorized as
\[\Big(L\,D_w+M\,{\mb I}\Big)\,D_w\,F=0,\]
with obvious $L$ and $M$. This factorization leads to
\beq\label{ng1}
F_1=1\qq\qq F_2=\int\, e^{-U(w)}dw,\qq 
U(w)=\int\,\frac{dw}{w^{\ga}(1-w)^{\de}(1-k^2 w)^{\eps}}.\eeq
Starting from these solutions we can use the change of function
\beq\label{Fhom}
F\to \wti{F}:\qq F(w)=w^{\rho}(1-w)^{\si}(1-k^2 w)^{\tau}\wti{F}(w)\eeq
which transforms, for special values of the parameters $\rho,\,\si,\,\tau$, a Heun 
function into another Heun function, up to changes in the parameters \cite{Ar}[p.18]. 
In such a way one generates 7 more Heun functions starting from (\ref{ng1}). For all 
these solutions Heun's differential operator remains obviously factorized 
and,  as shown in \cite{Ron2}, this happens only for these cases.

\section{Derivatives of Heun functions}
In general the derivative of a Heun function cannot be expressed in terms 
of some Heun function with different parameters. However, as shown in \cite{IK}, this 
happens in 4 cases:
\beq\label{dh1}
\barr{l}\dst
Hn\,'(k^2,0;\alf,\be,\ga,\de;w)=-\frac{\alf\be}{\ga+2}k^2w\,
Hn(k^2,s';\alf+2,\be+2,\ga+2,\de+1;w),\\[4mm]
-s'=\ga+\de+1+(\ga+\eps+1)k^2.\earr
\eeq
\beq\label{dh2}
\barr{l}\dst
Hn\,'(k^2,-\alf\be k^2;\alf,\be,\ga,\de;w)=\frac{\alf\be}{\ga+1}k^2(1-w)\,
Hn(k^2,s';\alf+2,\be+2,\ga+1,\de+2;w),\\[4mm]
-s'=\ga+\de+1+(\ga+\eps+\alf\be)k^2.\earr
\eeq
\beq\label{dh3}
\barr{l}\dst
Hn\,'(k^2,-\alf\be;\alf,\be,\ga,\de;w)=\frac{\alf\be}{\ga+1}(1-k^2w)\,
Hn(k^2,s';\alf+2,\be+2,\ga+1,\de+1;w),\\[4mm]
-s'=\ga+\de+\alf\be+(\ga+\eps+1)k^2.\earr
\eeq
\beq\label{dh4}
\barr{l}\dst
Hn\,'(k^2,s;0,\be,\ga,\de;w)=-\frac{s}{\ga+1}\,
Hn(k^2,s';2,\be+1,\ga+1,\de+1;w),\\[4mm]
s'=s-\ga-\de-(\ga+\eps)k^2.\earr
\eeq
Notice that, from (\ref{oph3}), the functions $Hn(\cdots;w)$ are analytic 
around $w=0.$

\section{Reduction to hypergeometric functions}
For non-generic solutions it was realized some time ago the possibility 
for Heun functions to reduce to hypergeometric functions. Some relations, using 
Weierstrass elliptic functions are given in \cite{Ka}.  
Later on Kuiken \cite{Ku} has observed some reductions to hypergeometric 
functions of some particular rational variable $R(w)$. This may happen 
only for {\em polynomial} $R(w)$, with the following list :
\beq\label{kk1}\barr{l}\dst 
{\rm Hn}\,(-1,0;2a,2b,2c-1,1+a+b-c;w)
=\,  _2F_1\left(\barr{c} a,b \\ c\earr \ ; w^2\right),\\[4mm]\dst 
{\rm Hn}\,(1/2,-2ab;2a,2b,c,1+2(a+b-c);w)
=\,  _2F_1\left(\barr{c} a,b \\ c\earr \ ; w(2-w)\right),\\[4mm]\dst 
{\rm Hn}\,(2,-4ab;2a,2b,c,c;w)
=\,  _2F_1\left(\barr{c} a,b \\ c\earr \ ; 4w(1-w)\right).\earr
\eeq
Notice that in all these cases $R(w)$ is a second degree polynomial with 3 free 
parameters. These results have been 
completed recently by Maier \cite{Ma1}, who discovered new cases \footnote{We have 
omitted the cases involving complex values of $k^2.$} with cubic and quartic 
polynomials $R(w)$ and 2 free parameters:
\beq\label{kk2}\barr{l}\dst 
{\rm Hn}\,(1/4,-9ab/4;3a,3b,1/2,2(a+b);w)
=\,  _2F_1\left(\barr{c} a,b \\ 1/2\earr \ ; w(3-w)^2/4\right),\\[4mm]\dst 
{\rm Hn}\,(1/2,-8ab;4a,4b,a+b+1/2,2(a+b);w)=\\[4mm]\dst 
\hspace{6cm}\,  _2F_1\left( \barr{c} a,b \\ a+b+1/2\earr \ ; 4w(2-w)(1-w)^2\right).\earr
\eeq
These results are not the whole story, since Heun functions may reduce to the 
product of some function $f(w)$ by some hypergeometric function with variable $R(w),$ 
not necessary polynomial. A result of this kind was obtained in \cite{Jo} by Joyce :
\beq\label{jo}
{\rm Hn}\,(1/4,-1/8;1/2,1/2,1,1/2;w)=\sqrt{\sqrt{4-w}-\sqrt{1-w}}\,
_2F_1\left(\barr{c} 1/2,1/2 \\ 1\earr \ ;R(w)\right),\eeq
with 
\[R(w)=\frac 14\,\left(\rule{0mm}{4mm}
2-w\sqrt{4-w}-(2-w)\sqrt{1-w}\right),\qq w\in\,[0,1].\]

Let us turn ourselves to the generic solutions of Heun's equation.

 \[  \]

\centerline{\bf\Huge Generic solutions}

\[  \]

\setcounter{section}{0}
\section{The 192 solutions of Heun's equation}
As Heun himself observed \cite{He} there is  a set of 24 substitutions of the variable 
$w$ which produce a transformation of Heun's equation into another Heun's equation with 
different parameters. This leads to a complete list of 192 solutions. This list has been 
fully worked out recently by Maier in \cite{Ma2}. We will just quote the 
generalizations of the Euler transformation of the hypergeometric function:
\beq\label{ma1}\barr{l}
Hn(k^2,s;\alf,\be;\ga,\de,\eps;w)=\\[4mm]
\hspace{2cm}=(1-w)^{1-\de}
Hn(k^2,-s-\ga(\de-1);\alf-\de+1,\be-\de+1;\ga,2-\de;w),\earr
\eeq
and of the Pfaff transformation
\beq\barr{l}
Hn(k^2,s;\alf,\be;\ga,\de,\eps;w)=\\[4mm]
=(1-w)^{-\alf}\,
Hn(-k^2/k'\,^2,-k^2/k'\,^2(s+\alf\ga);\alf,\alf-\de+1,\ga,\alf-\be+1;w/(w-1)).\earr
\eeq

\section{An integral transform}
An integral transform was given in \cite{Va1}:
\beq\label{it}\barr{l}
Hn(k^2,s;\alf,\be,\ga,\de;w)=\\[4mm]\dst 
= \frac{\G(\ga)}{\G(\alf)\G(\ga-\alf)}\,\int_0^1\,
t^{\alf-1}(1-t)^{\ga-\alf-1}\,Hn(k^2,s;\ga,\be,\alf,\de+\ga-\alf;wt)\,dt,\earr
\eeq
valid for ${\rm Re}\,\ga>{\rm Re}\,\alf>0$ and $w\in{\mb C}\backslash\,[1,\nf[.$

Let us use orthogonal polynomials to prove this relation \cite{Va2}. To this end 
we define the monic polynomials $M_n$ by
\[M_0=F_0=1,\qq M_n(P;x)=\mu_1\cdots\mu_n\,F_n(P;x)=n!\,(\ga)_n\,F_n(P;x),
\quad n\geq 1,\]
where $P$ denotes the set of parameters $\,k^2,\,\alf,\,\be,\,\ga,\,\de.$ These 
monic polynomials satisfy the recurrence relation:
\[\barr{l}
(\la_n+\mu_n+\ga_n-x)M_n=M_{n+1}+\la_{n-1}\mu_n\,M_{n-1},\quad n\geq 0,\\[4mm]
 M_{-1}(x)=0,\quad M_0(x)=1.\earr\]
If we define the new set of parameters 
\[P'=(k^2,\,\alf'=\ga,\,\be'=\be,\,\ga'=\alf,\,\de'=\de+\ga-\alf),\qq 
x'=s+\alf'\be'k^2\] 
it is easy to check the invariance relation $M_n(P';x')=M_n(P;x).$ This induces  
\beq\label{inv}
\quad x=s+\alf\be k^2,\qq\quad F_n(P;x)= \frac{(\alf)_n}{(\ga)_n}\,F_n(P';x'),\qq\quad 
x'=s+\alf'\be' k^2.\eeq
For ${\rm Re}\,\ga>{\rm Re}\,\alf>0$ we can write
\[\frac{(\alf)_n}{(\ga)_n}= \frac{\G(\ga)}{\G(\alf)\G(\ga-\alf)}\,
\int_0^1\,t^{n+\alf-1}(1-t)^{\ga-\alf-1},\]
then multiplying both sides of this relation by (\ref{inv}) and summing for $n\geq 0$ 
gives (\ref{it}) for $|w|<1.$ Analytic continuation extends it to 
$w\in{\mb C}\backslash\,[1,\nf[.$

\vspace{4mm}
\noindent{\bf Remarks :}

\brm
\item This is not an integral equation, since the parameters of the Heun function are 
changed in the transformation. Notice that several integral equations are known 
\cite{Ar}.
\item For $k^2\to 0$ we recover Bateman's integral relation 
\beq\label{bat}
_2F_1\left(\barr{c} a,b \\ c\earr \ ; w\right)=\frac{\G(c)}{\G(\la)\G(c-\la)}\,
\int_0^1\,t^{\la-1}(1-t)^{c-\la-1}\ 
_2F_1\left(\barr{c} a,b \\ \la\earr \ ; wt\right)\,dt, 
\eeq
valid for ${\rm Re}\,c>{\rm Re}\,\la>0.$ Notice that now $\la$ is a {\em free} 
parameter, and this enables one to deduce Euler's integral 
representation. This does not work for Heun functions because the 
parameter $\alf$ is not free.
\item For $k^2=1$ we get the relation
\beq\barr{l}\dst 
(1-w)^r\, _2F_1 \left(\barr{c} r+\alf,r+\be\\ \ga\earr \ ; w\right)=\\[4mm]\dst 
=\frac{\G(\ga)}{\G(\alf)\G(\ga-\alf)}\,\int_0^1\,
t^{\alf-1}(1-t)^{\ga-\alf-1}\,(1-wt)^{\tilde{r}}\, 
_2F_1 \left(\barr{c} \tilde{r} +\ga,\tilde{r}+\be\\ \alf\earr \ ; wt\right)\,dt,\earr
\eeq
with
\[r=\rho-\sqrt{\rho^2-\alf\be-s},\quad \rho=\frac 12(\ga-\alf-\be),\quad  
\tilde{r}=\tilde{\rho}-\sqrt{\tilde{\rho}\,^2-\be\ga-s},\quad 
\tilde{\rho}=-\frac 12(\ga-\alf+\be).\]
This relation does not appear in the extensive list of hypergeometric integrals given 
in \cite{ET}, so it could be new.
\erm

\section{Carlitz solutions}
In his analysis of some orthogonal polynomials of Stieltjes, Carlitz \cite{Ca} 
discovered the following remarkable result: the linearly independent Heun 
functions with parameters$\,(k^2,s\neq 0;0,1/2,1/2,1/2)\,$ are given by
\beq\label{ca1}
\exp\big(\pm 2i\,\sqrt{s}\,z(w)\big),\qq 
z(w)=\int_0^w\frac{du}{2\sqrt{u(1-u)(1-k^2u)}}.
\eeq
To check most simply this result we note that by the inversion theorem of 
elliptic functions we can write
\[
z(w,k^2)=\int_0^w\frac{du}{2\sqrt{u(1-u)(1-k^2u)}}\quad\Longleftrightarrow\quad 
w=\sn^2 (z,k^2).
\]
This conformal transformation maps the singular points as
\[\barr{ccccc} w\qq & 0\qq & 1\qq &\dst\frac 1{k^2}\qq &\nf\\[6mm]
\downarrow\qq &\downarrow\qq & \downarrow\qq & \downarrow\qq & \downarrow \\[6mm]
z\qq & 0\qq & K\qq & K+iK'\qq & iK'\earr
\]
where $K=K(k^2)$ and $K'=K(k'\,^2)$ are the complete elliptic integrals of first kind.
Using $z$ as new variable and setting $F(w)=y(z)$, Heun's equation becomes
\beq\label{hl1} \barr{l}\dst 
\frac{d^2y}{dz^2}+\left[(2\ga-1)\frac{\cn z\,\dn z}{\sn z}
-(2\de-1)\frac{\sn z\,\dn z}{\cn z}-(2\eps-1)k^2\frac{\sn z\,\cn z}{\dn z}
\right]\frac{dy}{dz}\\[5mm]\dst 
\hspace{4cm}+4(s+\alf\be\, k^2\,\sn^2 z)y=0.\earr
\eeq
At the very symmetric point where $\alf\be=0$ and $\ga=\de=\eps=1/2,$ it reduces to
\[\frac{d^2y}{dz^2}+4s\,y=0,\]
which proves (\ref{ca1}).
 
Using then the transformation (\ref{Fhom}) we can generate a full set of 8 generic 
solutions (see \cite{Va1}) where the detailed list is given). So we realize that 
Heun's equation, in the elliptic functions setting, lies in the field of differential 
equations with doubly periodic coefficients. It happens that 
Floquet's theory for differential equations with periodic coefficients does 
generalize to the case of doubly periodic coefficients and was derived by Picard.  
This will be discussed in the next subsection.

\section{Elliptic functions of second kind and Picard's theorem}
Since this theorem is not easily available in the standard textbooks 
\cite{Ak2},\cite{WW}, we will present some background material. Let us just 
recall the definition and a few basic results from elliptic function theory : 

\begin{ndef}
A function $\Phi(z)$ is elliptic if it is meromorphic and it has two 
linearly independent periods 
\[\Phi(z+2\om)=\Phi(z),\qq\qq \Phi(z+2\om')=\Phi(z).\]
These two periods define therefore a non-degenerate period parallelogram (usually 
$\,\om=K,\ \om'=iK'$).
\end{ndef}
We will need also
\begin{nth}
An elliptic function:
\brm
\item Has as many poles as zeroes in a period parallelogram.
\item Can be written
\beq\label{ell1}
\Phi(z)=A\frac{H(z-b_1)\cdots H(z-b_n)}{H(z-a_1)\cdots H(z-a_n)},\qq
a_1+\cdots +a_n=b_1+\cdots +b_n,\eeq
where $H(z)$ is one of Jacobi's theta function defined as
\[H(z)=2\sum_{n\geq 0}(-1)^nq^{(n+\frac 12)^2}\sin\left((2n+1)\frac{\pi z}{2K}\right),
\qq q=e^{-\frac K{K'}}.\]
\item Must have at least a pole of multiplicity 2 in a period 
parallelogram otherwise it is a constant.
\item With poles of multiplicity $n$ at $z=a$ can be expanded as:
\beq\label{ell2}
\Phi(z)=\sum a_0\,g(z-a)+a_1\,g'(z-a)+\cdots a_n\,g^{(n)}(z-a),\qq\quad 
g(z)=\frac{H'(z)}{H(z)},\eeq
where the sum is extended to all the poles.
\erm
\end{nth}

To state Picard's theorem we need the definition

\begin{ndef}
A function $\Phi(z)$ is a an elliptic function of second kind (or a function with 
constant multipliers) if it is meromorphic and that there exists two non-vanishing 
constants $\mu$ and $\mu'$ such that
\[\Phi(z+2\om)=e^{\mu}\,\Phi(z)\qq\quad \Phi(z+2\om')=e^{\mu'}\,\Phi(z),\]
where $2\om$ and $2\om'$ are two linearly independent periods. The constants $\,\mu\,$ 
and $\,\mu'\,$ are called the multipliers.
\end{ndef}

Obviously for $e^{\mu}=e^{\mu'}=1$ we recover elliptic functions. We will have to 
discuss separately the two cases:
\brm
\item Generic multipliers for which $\,\om\mu'-\om'\mu\neq 0$,
\item Special multipliers for which $\,\om\mu'-\om'\mu= 0$.
\erm

\subsection{Theorems for generic multipliers}
Let us consider the function
\beq\label{gm1}
f(z)=e^{\la z}\frac{H(z-a)}{H(z)},\eeq
where $a$ is not homologous to zero. One can check that it is elliptic of second kind 
with generic multipliers
\[\mu=2\la K,\qq\quad \mu'=i\left(2\la K'-i\pi \frac aK\right).\]
Let us prove:

\begin{nth}An elliptic function of second kind with generic multipliers:
\brm
\item Admits as many zeroes as poles in a period parallelogram.
\item Must have at least a simple pole in a period parallelogram, otherwise it vanishes.
\item Can be expanded, using $f(z)$ defined in (\ref{gm1}), as
\beq\label{gm2}
\Phi(z)= \sum c_0f(z-a)+c_1f'(z-a)+\cdots+c_nf^{(n)}(z-a),
\eeq
where the sum extends to all poles $z=a$ of multiplicity $n$ of $\Phi.$
\erm
\end{nth}

\nin{\bf Proof :} 

\nin Notice that for any elliptic function of second kind $\Phi(z)$ with given generic 
multipliers $\,(\mu,\mu')\,$ it is always possible to 
find $\la$ and $a$ in (\ref{gm1}) such that $f$ has the {\em same} multipliers. It 
follows that $\Phi(z)/f(z)$ is elliptic. Using (\ref{ell1}) we 
get the most general structure
\[ \Phi(z)=A\,f(z)\,\frac{H(z-b_1)\cdots H(z-b_n)}{H(z-a_1)\cdots H(z-a_n)},\qq
a_1+\cdots +a_n=b_1+\cdots +b_n,\]
which proves the first assertion.

For the second assertion, let it be supposed that $\Phi$ has no pole: it cannot 
have any zero and therefore $\Phi(z)$ reduces to $Ae^{\la z},$ but in this case the 
multipliers are not generic, contradicting our hypothesis, hence $A=0.$ 

For the third assertion, let us consider a pole $z=a$ of multiplicity $n.$  
Near to $z=a$ we have the Laurent expansion
\[f^{(k)}(z)=\frac{\xi_k}{(z-a)^{k+1}}+{\rm holomorphic},\]
so we can find coefficients $\{c_k,\,k=0\ldots n\}$ such that 
\[\Phi(z)-\sum (c_0f(z-a)+c_1f'(z-a)+\cdots+c_nf^{(n)}(z-a))\]
has no poles in a period parallelogram, so it must vanish. $\qq\Box$

\subsection{Theorems for special multipliers}
This time let us define
\beq\label{sm0}
f(z)=e^{\la z}\frac{H'}{H}(z).
\eeq
One can check that it is elliptic of second kind 
with special multipliers. We have now

\begin{nth}
Any elliptic function of second kind with special multipliers can be expanded as 
\beq\label{sm1}
\Phi(z)=C\,e^{\la z}+\sum\left(c_0f(z-a)+c_1f'(z-a)+\cdots c_n\,f^{(n)}(z-a)\right),
\eeq
with the constraint
\beq\label{sm2}
\sum e^{-\la a}(c_0+c_1\,\la+\cdots c_n\,\la^n)=0,
\eeq
where the sums extend to all poles of $\Phi(z).$
\end{nth}

{\nin {\bf Proof:}

\nin In this case $\om\,\mu'-\om'\,\mu=0.$ So we can find a value of $\la$ such that 
$F(z)\equiv \Phi(z)e^{-\la z}$ is elliptic. Let us consider a pole $z=a$ of 
order $n$ of $\Phi.$ We have for Laurent's series
\[\Phi(z)=\frac{c_0}{z-a}-\frac{c_1}{(z-a)^2}+\cdots
+(-1)^n \frac{n!\,c_n}{(z-a)^{n+1}}+{\rm holomorphic}.\]
hence we can write
\[F(z)=\frac{a_0}{z-a}-\frac{a_1}{(z-a)^2}+\cdots
+(-1)^n\frac{n!\,a_n}{(z-a)^{n+1}}+{\rm holomorphic},\]
where the new residue is
\[a_0=e^{-\la a}(c_0+\la c_1+\cdots+\la^n c_n).\]
Using the expansion theorem for elliptic functions (\ref{ell2}) we have
\[
\barr{l}\dst
\Phi(z)e^{-\la z}=C+\sum e^{-\la a}\left[\rule{0mm}{4mm}
a_0\,\frac{H'}{H}(z-a)+a_1\,D_z\left(\frac{H'}{H}(z-a)\right)+\cdots\right.\\[4mm]\dst
\left.\hspace{8cm}\cdots+a_n\,D_z^{n}\left(\frac{H'}{H}(z-a)\right)\rule{0mm}{4mm}
\right].\earr\]
Inserting $e^{\la z}$ into the right-hand side, and expanding the derivatives 
according to Leibnitz rule gives (\ref{sm1}). The constraint (\ref{sm2}) 
comes from the fact that the sum of the residues in a period parallelogram vanishes 
for an  elliptic function. $\qq\Box$

\subsection{Picard's theorem}
Now we can state Picard's theorem \cite{Pi1}

\begin{nth}
Let us consider the differential equation
\[\frac{d^nF}{dz^n}+a_1(z)\frac{d^{n-1}F}{dz^{n-1}}+\cdots a_n(z)F=0,\]
with doubly periodic coefficients 
\[a_l(z+2\om)=a_l(z),\qq a_l(z+2\om')=a_l(z),\qq l=1,2,\ldots,n,\]
Any {\em meromorphic} solution $F(z)$ is a linear combination of elliptic 
functions of the second kind. 
\end{nth}

\nin{\bf Proof :}

\nin To shorten, and in view of application to Heun's case, we will consider 
a differential equation of second order: 
\beq\label{pic1}
\frac{d^2F}{dz^2}+p(z)\frac{dF}{dz}+q(z)F=0,\eeq
with doubly periodic coefficients $p$ and $q.$ Let us consider a solution $F(z)$ 
which is not elliptic of second kind: then the ratio $F(z+2\om)/F(z)$ cannot 
be a constant. The periodicity of the coefficients implies that the functions 
$F(z+2\om)$ and $F(z+4\om)$ are also solutions, so we must have a linear 
relation of the form
\[F(z+4\om)=AF(z)+BF(z+2\om).\]
Let us now consider the non-vanishing function $\phi(z)=F(z+2\om)+\rho F(z).$ 
If we take for $\rho$ a root of $\rho^2+B\rho-A=0$ it is easy to check that 
$\phi(z+2\om)=(B+\rho)\phi(z).$ 

A similar argument works with respect ot the period $2\om'.$ So we have 
proved that we can construct a first solution $\phi(z)$ which is elliptic of 
the second kind.

We have now to prove that a second linearly independent and meromorphic solution, 
which can be written as
\[\psi(z)=\phi(z)\int G(z)\,dz,\qq G=\frac 1{\phi^2(z)}\,e^{-P},\qq P'=p,\]
is also elliptic of second kind.

The Wronskian
\[\psi\,\phi'-\psi'\,\phi=C\,e^{-P},\qq C\neq 0,\]
and the meromorphy of $\phi$ and $\psi$ imply that $e^{-P}$ is meromorphic. Since 
$p$ is doubly periodic, $e^{-P}$ is elliptic of second kind. So $G$ is a 
meromorphic elliptic function of second kind. 

Let it be supposed first that $G$ has generic multipliers. Using the expansion 
theorem (\ref{gm2}) we can write
\[G(z)= \sum\left[\rule{0mm}{4mm}c_0\,f(z-a)+\cdots
+c_n\,f^{(n)}(z-a)\right],\qq f(z)=e^{\la z}\frac{H(z-a)}{H(z)},\]
where the sum includes all the poles of multiplicity $n$ of $G(z).$ 
Since the integral has to be meromorphic all the coefficients $c_0$ must vanish 
and we get
\[{\cal G}(z)\equiv\int G(z)\,dz= \sum\left[\rule{0mm}{4mm}c_1\,f(z-a)+\cdots
+c_n\,f^{(n-1)}(z-a)\right],\]
and from (\ref{gm2}) it follows that $\,{\cal G}(z)$ and $\psi(z)$ are 
elliptic of second kind. 

Let us now consider the case where $G$ has special multipliers. The expansion 
theorem (\ref{sm1}) gives
\[G(z)=C\,e^{hz}+\sum\left[\rule{0mm}{4mm}c_0\,f(z-a)+\cdots
+c_n\,f^{(n)}(z-a)\right],\qq f(z)=e^{\la z}\frac{H'}{H}(z), \]
with the constraint
\beq\label{cont}
\sum\left(\rule{0mm}{4mm}c_0+c_1 h+\cdots+c_n h^n\right)e^{-ha}=0.\eeq

If $h\neq 0$ the meromorphy of $G$ requires that all coefficients $c_0$ vanish, so 
we can write
\[{\cal G}(z)=\frac Ch\,e^{hz}+
\sum\left[\rule{0mm}{4mm}c_1\,f(z-a)+\cdots+c_n\,f^{(n-1)}(z-a)\right].\]
This relation and the constraint (\ref{cont}), with coefficients $c_0$ all vanishing, 
implies that $\,{\cal G}(z)\,$ is elliptic of second kind with special multipliers, 
hence $\psi$ is again elliptic of second kind.

If $h=0$ we have
\[{\cal G}(z)=Cz+\sum\left[\rule{0mm}{4mm}c_1\,f(z-a)+\cdots
+c_n\,f^{(n-1)}(z-a)\right],\qq\sum\,c_0=0.\]
This implies
\[{\cal G}(z+2\om)={\cal G}(z)+D,\qq\quad {\cal G}(z+2\om')={\cal G}(z)+D'\]
and so, if $(\la_1,\la_2)$ are the multipliers of $\phi$ we can write
\[\psi(z+2\om)=\la_1\,\psi(z)+\la_1 D\,\phi(z),\qq\quad 
\psi(z+2\om')=\la_2\,\psi(z)+\la_2 D'\,\phi(z),\]
so we can subtract from $\,\psi\,$ a suitable term linear in $\phi$ for which 
$\psi$ will be elliptic of second kind.$\qq\Box$

\section{The meromorphic solutions}
So let us look for the necessary conditions on the parameters to get meromorphic 
solutions in the variable $z$. The computation of the exponents at the 
singularities of (\ref{hl1}) is quite simple and gives 
\beq\barr{lcl}
z=0\qq & (\sn z)^{2\rho_1}\qq & \quad (0,\ga-1)\\[4mm]
z=K\qq & (\cn z)^{2\rho_2}\qq & \quad (0,\de-1)\\[4mm]
z=K+iK'\qq & (\dn z)^{2\rho_3}\qq & \quad  (0,\eps-1)\\[4mm]
z=iK'\qq & (\sn z)^{-2\rho_3}\qq & \quad (\alf,\be)\earr
\eeq
So the necessary conditions for meromorphy are
\beq
\left\{\barr{l}
\ga=\frac 12-m_1,\qq \de=\frac 12-m_2,\qq \eps=\frac 12-m_3,\qq M=m_1+m_2+m_3,\\[4mm]
\alf=-\frac 12(m_0+M),\qq \be=\frac 12(m_0-M+1),\qq N=m_0+M,\earr\right.
\eeq
with the vector $\,\ol{N}=(m_0,\,m_1,\,m_2,\,m_3)\in{\mb Z}^4.$ That these 
conditions are also sufficient is proved in \cite{GW}. From Picard's Theorem the 
solutions will be elliptic functions of the second kind. The differential 
equation becomes
\beq\label{newH} \barr{l}\dst 
\frac{d^2y}{dz^2}+2\left(-m_1\frac{\cn z\,\dn z}{\sn z}
+m_2\frac{\sn z\,\dn z}{\cn z}+m_3\,k^2\frac{\sn z\,\cn z}{\dn z}
\right)\frac{dy}{dz}\\[5mm]\dst 
\hspace{5cm}+(4s+N(N-2m_0-1)\,k^2\,\sn^2 z)y=0,\earr
\eeq
and for $\ol{N}=(n,\,0,\,0,\,0)$ we are back to Lam\'e's equation.

It is possible, extracting from $y$ suitable factors of $w,$ to relate the negative 
and positive values of the parameters $m_i$ as summarized in \cite{Sm}[p. 296], 
so from now on we will consider $\,\ol{N}\in{\mb N}^4.$

It is interesting to get rid of the derivative in (\ref{newH}) by the change of 
function
\beq\label{HD1}
y(z)=(\sn z)^{m_1}\,(\cn z)^{m_2}\,(\dn z)^{m_3}\,Y(z)\qq\Longrightarrow\qq 
\frac{d^2Y}{dz^2}=(V(z)-A)Y
\eeq
with 
\beq\label{HD2}
\barr{l} 
V(z)=\frac{m_1(m_1+1)}{\sn^2 z}+m_2(m_2+1)\frac{\dn^2 z}{\cn^2 z}
+m_3(m_3+1)\frac{k^2\,\cn^2 z}{\dn^2 z}+m_0(m_0+1)\,k^2\,\sn^2 z \\[4mm]
A=4s+(m_1+m_2)^2+k^2(m_1+m_3)^2.\earr
\eeq

This equation has been considered by Darboux \cite{Da}. In this short article 
(3 pages) he claims:
\brm
\item[$\bullet$] that the product of two solutions is a polynomial, which we denote by 
$\Psi_{g,N}(\si;w)$ of degree $N$ in $w=\sn^2 z$ and of degree $g$ in $\si=4s.$ The 
knowledge of this polynomial is of paramount importance as we will see later. Let 
us quote Darboux: ``Une fois le polyn\^ome $\Psi$ d\'etermin\'e, l'int\'egration 
s'ach\`eve, comme on sait, sans aucune difficult\'e." 
\item[$\bullet$] that for half-integer values of $\,m_1,\,m_2,\,m_3\,$ this 
equation can be integrated (for arbitrary $A$ i. e. for what we call generic solutions).
But neither detailed proofs nor the explicit forms of the solutions (even for the 
simpler case of integer $m_i$) were given. 
\erm

We would like to point out that even for Lam\'e's equation with half-integer values 
of $n$ only {\em non-generic} solutions, due to Halphen and Brioschi, are known, 
so that the claim of Darboux seems questionable. 

To come back to Lam\'e equation, let us mention that its solutions are meromorphic 
for integer $n$, and their general form, due to Halphen and Hermite, is given 
for instance in \cite{WW}[p. 570-575]. However, since one has to solve a set of 
$n$ linear equations, only for low values of $n$ everything can be made explicit.

Interestingly enough, this differential equation has also appeared in the quite 
different field of integrable models, particularly KdV equation. Then relation 
(\ref{HD1}) can be interpreted as a Schr\"odinger eigenvalue problem, with 
eigenvalue $A$ (equivalent to the auxiliary parameter $s$) and potential 
$V(z)$ given by (\ref{HD2}). 

The solutions 
corresponding to $m_i\in{\mb Z}$ are called ``finite-gap" solutions and 
sometimes $V(z)$ is called a Treibich-Verdier potential \cite{TV}, after 
their work on the subject. The name ``finite-gap" refers to the appearance of a 
finite number of energy bands in the Bloch spectrum, a phenomenon discovered 
a long time ago by Ince \cite{In}.

\section{Elliptic level one solutions}

\begin{nth}
The level one ($N=1$) solutions, with $\si=4s,$ are given by:
\beq\label{KH}\barr{llll}
{\rm parameter}\quad & {\rm solution}\ y(z)\quad & {\rm constraint}\quad & 
\Psi(\si,w)\\[4mm]
m_0=1 & \dst e^{zZ(\om)}\frac{H(z-\om)}{\T(z)}\qq & \dn^2 \om=\si-k^2 & 
\si+k^2\,w-1-k^2 \\[5mm]
m_1=1 & \dst e^{zZ(\om)}\frac{\T(z-\om)}{\T(z)}\qq & \dn^2 \om=\si+1&
\si\,w+1 \\[5mm]
m_2=1 & \dst e^{zZ(\om)}\frac{\T_1(z-\om)}{\T(z)}\qq & \dn^2 \om=\si+1-k^2
& \si(1-w)+1-k^2 \\[5mm]
m_3=1 & \dst e^{zZ(\om)}\frac{H_1(z-\om)}{\T(z)}\qq & \dn^2 \om=\si & 
\si(1-k^2\,w)-1+k^2 \earr
\eeq
with $\dst Z(z)=\frac{\T'}{\T}(z).$
\end{nth}

\nin{\bf Proof:}

\nin We will give the detailed proof for $m_1=1$, the other cases being 
analyzed similarly. So we start with
\beq\label{m1proof1}
\frac{d^2y}{dz^2}-2\frac{\cn z\,\dn z}{\sn z}\frac{dy}{dz}+\si\,y=0,
\eeq 
and we look for a solution $\dst y=e^{\mu z}\frac{H(z-\rho)}{\T(z)}.$ 
Taking derivatives we get
\[\frac{y'}{y}= \mu+\frac{H'}{H}(z-\rho)-\frac{\T'}{\T}(z),\qq  
\frac{y''}{y}= \left(\frac{y'}{y}\right)^2 -\frac 1{\sn^2(z-\rho)}+k^2\,\sn^2 z.\]
Let us define the auxiliary function
\[ \chi(z)\equiv \left(\frac{y'}{y}\right)^2 
-2\frac{\cn z\,\dn z}{\sn z}\,\frac{y'}{y}-\frac 1{\sn^2(z-\rho)}+k^2\,\sn^2 z+\si,\]
so that proving that $\chi$ vanishes identically is equivalent to proving that $y$ is 
indeed a solution of (\ref{m1proof1}).

For $z=-iK'$ one can check that the poles in $\chi$ cancel automatically. Imposing 
that the pole at $z=0$ is absent gives $\dst \mu=\frac{H'}{H}(\rho).$ The absence 
of the pole at $z=\rho$ is easily checked. So we know that $\chi$ is bounded in 
a period parallelogram and since it is doubly periodic, it is bounded in all the 
complex plane: by Liouville theorem it is a constant. Imposing 
that $\chi$ vanishes for $z=K$ fixes the value of $\rho$ by the equation 
$\sn^2\rho=-1/\si.$ Then we switch to the new parameter $\om=\rho+iK'$ and we express 
the solution in terms of $\om$ using the transformation theory of theta functions. 

We can now determine the spectral polynomial $\Psi(\si;z)$ 
defined, up to an overall constant factor, by the product of the two solutions of 
Heun's equation. Let us do it first for $m_0=1.$ The product of the two 
solutions is computed, using the transformation theory of theta functions, to 
the identity
\beq\label{theta}
\frac{H(z-\om)H(z+\om)}{\T^2(z)\T^2(\om)\T^2(0)}=k^2\,\sn^2 z-k^2\,\sn^2 \om,
\eeq 
which is a first degree polynomial with respect to $w=\sn^2 z$ and to 
$\si=1+k^2-k^2\sn^2\om.\qq\Box$

\vspace{4mm}
\nin{\bf Remarks:}

\brm
\item Notice that $a$ is uniquely defined up to congruence.
\item All the solutions have genus $g=1.$
\item The other linearly solution is obtained, for generic $s\neq 0$ by the change 
$z\to\ -z.$
\item The solution for $m_0=1$ is due to Hermite \cite{WW}[p.573] and the solution 
for $m_3=1$ is due to Picard \cite{Pi2}.
\item The first two formulas for $\Psi(\si;w)$ agree with the results given in \cite{Sm}.
\item For special values of $\si,$ or equivalently of $a,$ we may fail to get 
two linearly independent solutions. In this case factoring out the solution $y$ given 
previously gives the second solution via a quadrature. Let us give some 
examples for the solution with $m_0=1$:
\beq\barr{ll}
\si=1+k^2\qq & \dst A\,\sn z+B\,\sn z\left(\frac{H'}{H}(z)+\frac{E-K}{K}\,z\right)\\
[5mm]
\si=1\qq & \dst A\,\cn z
+B\,\cn z\left(\frac{H_1}{H_1}(z)+\frac{E-k'\,^2K}{K}\,z\right)\\[5mm]
\si=k^2\qq & \dst A\,\dn z
+B\,\dn z\left(\frac{\T'_1}{\T_1}(z)+\frac EK\,z\right).\earr
\eeq
\item The general structure of the elliptic solutions, for arbitrary level, are 
given in \cite{GW}. They suffer from the same defect as the general solution of 
Lam\'e's equation: they lead to really explicit expressions only for low values 
of the level $N.$
\erm

\section{Finite-gap solutions}
In \cite{Sm}, the following facts were used:
\brm
\item Let us call $\Psi(\si=4s;w)$ the product of two solutions of Heun's equation 
and let us define  $p$ (resp. $q$) as the coefficient of the 
derivative (resp. of the function) in (\ref{eqHn1}). Then $\Psi$ must be a solution 
of the third order differential equation
\beq\label{ed3}
\Psi'''+3p\Psi''+(p'+2p^2+4q)\Psi'+2(q'+2pq)\Psi=0.\eeq
For the meromorphic solutions of Heun's equations, with parameters
\[\left\{\barr{l}
\ga=\frac 12-m_1,\qq \de=\frac 12-m_2,\qq \eps=\frac 12-m_3,\qq M=m_1+m_2+m_3,\\[4mm]
\alf=-\frac 12(m_0+M),\qq \be=\frac 12(m_0-M+1),\qq N=m_0+M,\earr\right.\]
the product $\Psi_{g,N}(\si;w)$ is a polynomial of degree $N$ with respect 
to the variable $w$:
\beq\label{cspectr1} 
\Psi_{g,N}(\si,w)= a_0(\si)\,w^N+a_1(\si)\, w^{N-1}+\cdots+a_N(\si),\eeq
and a polynomial of degree $g$ in the parameter $\si$:
\beq\label{cspectr2}
\Psi_{g,N}(\si,w)=b_0(w)\,\si^g+b_1(w)\,\si^{g-1}+\cdots+b_g(w).
\eeq
The leading term $b_0$ is given by a first order differential equation due to the 
fact that $\si$ appears linearly only in the coefficients of $\Psi'$ and $\Psi$ 
of equation (\ref{ed3}). We have taken for normalization
\[b_0(w)=w^{m_1}(1-w)^{m_2}(1-k^2w)^{m_3},\]
as can be checked from (\ref{KH}).
\item Then, following a method due to Lindemann and Stieltjes \cite{WW}[p. 420] 
adapted to Heun's case, one looks for a solution of the form
\[F(w)=\sqrt{\Psi}
\exp\left(\pm i\frac{\nu(\si)}{2}\int\frac{N(w)}{\Psi(w)}\, dw\right).\]
Inserting this ansatz into Heun's equation gives on the one hand
\[\frac{N'}{N}=-p\qq\Longrightarrow\qq N(w)=\frac{w^{m_1}(1-w)^{m_2}(1-k^2 w)^{m_3}}
{\sqrt{w(1-w)(1-k^2 w)}}\]
(recall that all the $m_i$ are positive) and on the other hand:
\beq\label{nu1}
\nu^2(\si)=\frac{2\Psi\Psi''-\Psi'\,^2+2p\Psi\Psi'+4q\Psi^2}{N^2},
\eeq
showing that $\nu^2$ is of degree $2g+1$ in $\si.$ The fact that $\nu$ is a constant 
is easily checked by differentiating relation (\ref{nu1}) and using the 
differential equation (\ref{ed3}).

So we conclude that Heun's functions (the so-called ``finite-gap" solutions), 
with the parameters given above, have for integral representation: 
\beq\label{fg1}
\sqrt{\Psi_{g,N}(\la;w)}\,\exp\left(\pm i\frac{\nu(\la)}{2}
\int\frac{w^{m_1}(w-1)^{m_2}(1-k^2w)^{m_3}\,dw}
{\Psi_{g,N}(\la;w)\sqrt{w(w-1)(1-k^2w)}}\right).
\eeq
\erm
For the level one solutions already considered the polynomial $\nu^2(\si)$ is 
given by
\beq\label{nu2}
\barr{ll}
m_0=1\qq\quad & (\si-1)(\si-k^2)(\si-1-k^2)\\[4mm]
m_1=1     & \si(\si+1)(\si+k^2)\\[4mm]
m_2=1     &\si(\si-k^2)(\si+1-k^2)\\[4mm]
m_3=1     &\si(\si-1)(\si-1+k^2)\earr
\eeq

A partial list of the finite-gap solutions, up to level $N=5$, has been 
given in \cite{Sm}.

These results show that there do exist integral representations for Heun functions, 
but they are not ``cheap". Furthermore, as we will show now on a particular example, 
these integral representations  are just a different dressing of the elliptic 
solutions: they are a kind of ``algebraization" of the elliptic solutions, but are 
exactly the same analytic objects.

\section{Finite-gap versus elliptic solutions}
We will show, for the level one finite-gap integral representation with 
$m_1=1,$ that it does coincide with its corresponding elliptic solution given 
by (\ref{KH}). The other cases can be relateded using completely similar arguments. 
For this identification it is sufficient to consider $w\in[0,1].$ We start 
from the data
\[ \si=-k^2\,\sn^2 \om,\qq \nu^2(\si)=\si(\si+1)(\si+k^2),\qq 
\Psi(\si;w)=\si\,w+1.\]
So the factor appearing in the exponential of relation (\ref{fg1}) is
\[\pm \frac 1{2} \ \int\,
\frac{k^2\,\sn \om\,\cn \om\,\dn \om\ w\,dw}{(1+\si\,w)\sqrt{w(1-w)(1-k^2\,w)}}.\]
Under the change of variable $w=\sn^2 z,$ with $z\in\,[0,K],$ it becomes an elliptic 
integral of the third kind, computed in \cite{WW}[p. 523]:
\[\pm\,\int\frac{k^2\sn \om\,\cn \om\,\dn \om\,\sn^2 z\,dz}{1-k^2\,\sn^2 \om\,\sn^2 z}
=\pm\frac 12\left(\ln\frac{\T(z-\om)}{\T(z+\om)}+z\,Z(\om)\right).\]
So, keeping the plus sign, we get for the exponential term in (\ref{fg1})
\beq\label{form1}
e^{zZ(\om)}\ \sqrt{\frac{\T(z-\om)}{\T(z+\om)} }.
\eeq
And we need to multiply this by the square root of
\[ \Psi\propto 1-k^2\,\sn^2 \om\,\sn^2 z=\sn^2 \om\left(\rule{0mm}{4mm}
 k^2\,\sn^2(\om-iK')-k^2\,\sn^2 z\right),\]
and upon use of the identity (\ref{theta}) we are first left with
\[\Psi\propto\frac{H(-\om+z+iK')\,H(-\om-z+iK')}{\T^2(z)},\]
and after use of the transformation theory for the theta functions we end up with
\beq\label{form2}
\Psi\propto \frac{\T(z+\om)\,\T(z-\om)}{\T^2(z)}.
\eeq
Gathering (\ref{form1}) and (\ref{form2}) we conclude that the finite-gap solution 
(\ref{fg1}) reduces to
\[e^{zZ(\om)}\frac{\T(z-\om)}{\T(z)},\] 
which is nothing that the elliptic solution given in (\ref{KH}).

\section{Conclusion}
The finite-gap solutions or their elliptic counterparts solve Heun's equation for all 
cases where these solutions are meromorphic functions of the variable $z.$ It is 
interesting to notice that these progresses have been evolving in relation with 
integrability considerations. It is quite difficult to say what new ideas 
will require the generalization of these results to cover the non-meromorphic solutions. 
These much more difficult problems are left for the future.

\end{document}